\documentstyle[psfig, twocolumn]{esa_sp_latex}

\begin{document}
%

\parindent 0pt
\parskip 10pt plus 1pt minus 1pt
\hoffset=-1.5truecm
\topmargin=-1.0cm
\textwidth 17.1truecm \columnsep 1truecm \columnseprule 0pt 

\title{\bf Long-Term Variability in Bright Hard X-ray Sources: 5+ Years of BATSE Data}

\author{
\bf{C.R. Robinson$^1$, B.A. Harmon$^2$, M.L. McCollough$^1$, W.S. Paciesas$^3$,} \\
\bf{M. Sahi$^1$, D.M. Scott$^1$, C.A. Wilson$^2$, S.N. Zhang$^1$, K.J. Deal$^3$} \vspace{2mm} \\
$^1$ Universities Space Research Association, Huntsville, AL 35806 U.S.A. \\
$^2$ NASA/Marshall Space Flight Center, Huntsville, AL 35812 U.S.A. \\
$^3$ Department of Physics, University of Alabama in Huntsville, Huntsville, AL 35899 U.S.A.}

\maketitle

\begin{abstract}

The operation of CGRO/BATSE continues to produce, after more than 5 years, a
valuable database for the study of long-term variability in bright
hard X-ray sources.  The all-sky capability of BATSE provides, using the
Earth occultation technique, up to $\approx$30 flux measurements
per day for each source.  The long BATSE baseline and the numerous
rising and setting occultation flux measurements allow searches
for periodic and quasi-periodic signals from hours to hundreds of
days.

We present initial results from our study of the hard X-ray
variability in 24 of the brightest BATSE sources.  Power density
spectra are computed for each source.  In addition, we present
profiles of the hard X-ray orbital modulations in 8 X-ray binaries
(Cen X-3, \mbox{Cyg X-1,} Cyg X-3, GX 301-2, Her X-1, OAO1657-415, Vela X-1
and 4U1700-37), several-hundred-day modulations in the amplitude
and width of the main high state in the 35-day cycle in Her X-1,
and variations in outburst durations and intensities in the
recurrent X-ray transients.


\end{abstract}

\section{INTRODUCTION}

The continued operation of the BATSE experiment on CGRO
provides a valuable resource for the study of variability
in hard X-ray sources.  The BATSE experiment's Large Area Detectors (LADs)
consist of eight separate NaI(Tl) scintillation detectors
positioned in an octahedral pattern to provide continuous
coverage over 4$\pi$ steradians (see Fishman et al. 1989).
The detectors are sensitive from 20-1800 keV.  Although
uncollimated, the large collecting area, 2025 $\rm{cm^2}$ each, and 
the low-Earth orbit of CGRO provide the opportunity to monitor
various discrete X-ray objects using occultations by the Earth
(see Harmon et al. 1992).  Continuous monitoring also
allows the study of very bright and pulsed sources on time scales
shorter than the CGRO orbit (91-94 min) and the 8 Spectroscopy Detectors
are available for coverage down to around 8 keV, but we limit this analysis to
LAD occultation measurements of bright sources.
The ability of BATSE to provide essential information to \mbox{INTEGRAL}
on variable sources is discussed elsewhere in these proceedings
(see Fishman et al. 1997).

\section{ANALYSIS}
Flux measurements may be performed on individual occulted sources twice during each CGRO
orbit, at the times of source ingress and egress from the Earth's limb.
The effects of the Earth's atmosphere and detector background are deconvolved with
the resulting step down (ingress) or step up (egress) in the detected emission
providing
flux measurements in different hard X-ray energy bands.  Objects within around $\rm{45^{\circ}}$ of
the Celestial Equator are 
occulted during every CGRO orbit.  Sources closer to the Celestial Poles
are occulted during $\approx$80$\%$ or more of CGRO orbits
(modulated over the 49-52 day CGRO precession period).
This has produced up to 60,000 flux measurements per source.
Primarily due to source confusion and SAA passage, useful occultation
measurements are constrained to around 50$\%$ of this value.
Nevertheless, this still provides around 30,000 measurements per source allowing
variability studies on time scales from hours to several hundred days.

To search for periodic and quasi-periodic features,
power density spectra (PDS) were computed for 24 of the brightest hard X-ray sources
using data obtained from over 2000 days of Earth occultations (see Figures 1 and 2).
The data were extracted over different energy ranges, chosen to maximize signal-to-noise
(except for Cyg X-1 and the Crab where the energy ranges were chosen to be consistent with past analyses of these sources).
The unevenly sampled PDS were calculated according to the prescription of Scargle (1982).
Intrinsic source red noise, the CGRO precession period, times of pointing changes
(approximately every 1-2 weeks resulting in changes in detectors and detector angles to
sources), and variations in the background (most noticeable as diurnal modulations) provide
challenges to the extraction of true periodic signals from the celestial sources.   

While the plotted periodograms are normalized to the total
variance in the data (see Horne \& Baliunas 1986), we further normalized
the power spectra by the local mean power when determining the significance of
peaks and limits on sinusoidal modulations.  The plots exhibit
excess power at low frequencies produced by inherent red noise, in
sources like Cyg X-1, and power introduced through detector pointing
changes and periodic source interference.
In weaker sources located where the Earth's limb
will cross close to bright sources, the CGRO precession period is evident
near 0.02 cycles $\rm{d^{-1}}$
(e.g., Cen A and 4U 1608-522).  The changing precession
period, with spacecraft altitude, and beats between other signals broaden
the precession period peak and produce additional peaks.  Of more importance
presently, but with improvements still possible, is the effect due to
pointing changes.  Each flux measurement consists of contributions of
up to 4 LADs with each detector at a different
angle to the source.  Small variations in the response of each detector
and deviations from a cosine in the angular response,
particularly at near normal incidence to the detectors, produce small
discontinuities in the measured flux at
the boundaries between pointing changes.  In the Crab Nebula, for example, the
flux measurements just prior to TJD 10200 show an increase attributable
to the increased sensitivity in one of the detectors when the source was at
near normal incidence.

\begin{figure*}
\hbox{
\psfig{file=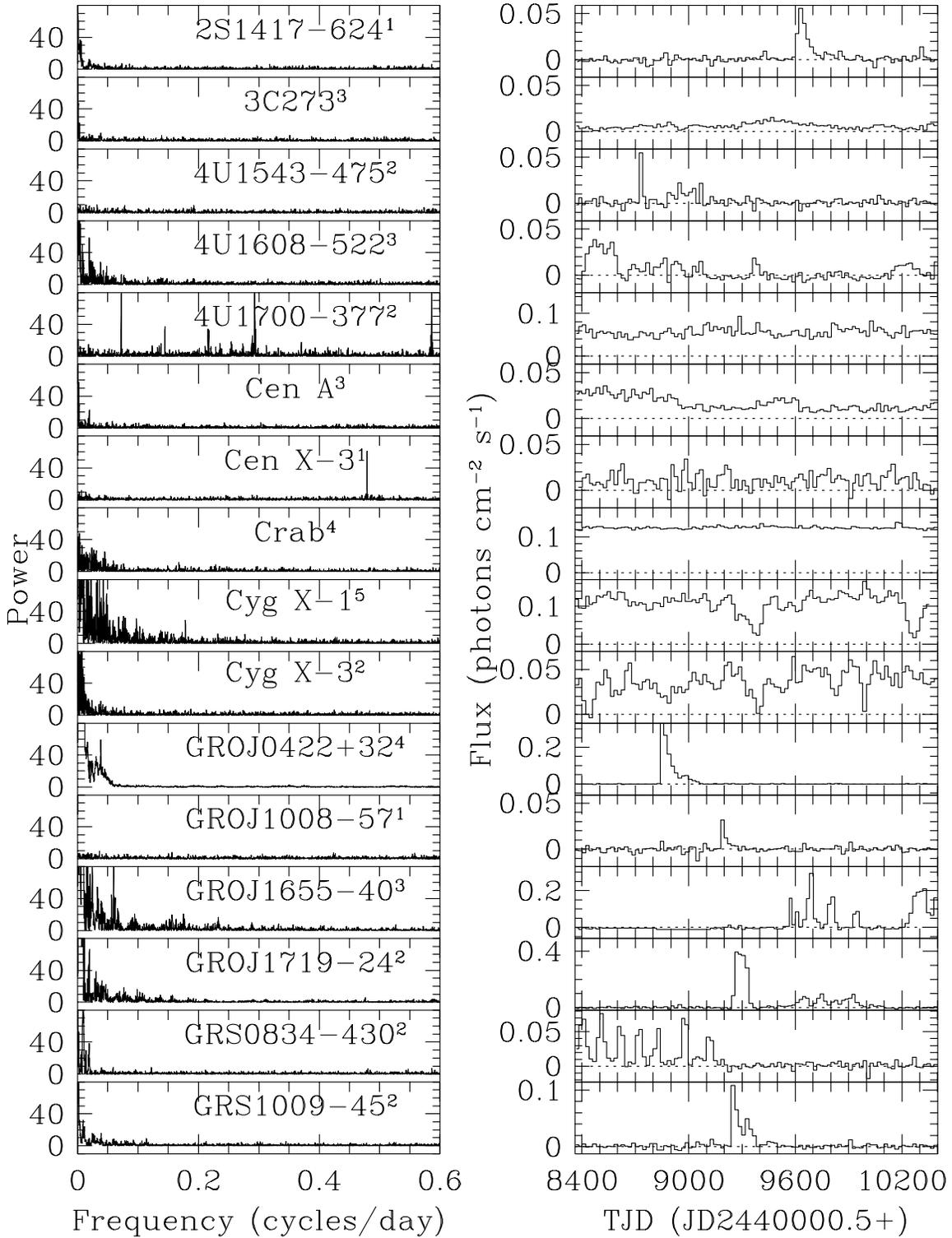,width=6.5in}
     }
\caption{ Power density spectra (left-hand side, computed to the ``mean'' Nyquist frequency ($\approx$8 cycles $\rm{d^{-1}}$) but
plotted only up to 0.6 cycles $\rm{d^{-1}}$) and their corresponding X-ray light curves (right-hand side, 20-day bins) are
shown for bright hard X-ray sources in the sample.  The superscripts following the source names refer to the
energy band chosen for analysis and plotting: 1 = 20-50 keV, 2 = 20-100 keV, 3 = 20-200 keV,
4 = 40-150 keV and 5 = 45-140 keV.}
\end{figure*}

\section{RESULTS}

\subsection{Orbital Modulation}
Hard X-ray eclipses and modulations at the system orbital periods were detected
for 8 X-ray binaries in our sample: 4U1700-377 (3.4d),
Cen X-3 (2.0d), Cyg X-1 (5.6d), Cyg X-3 (0.2d), \mbox{GX 301-2 (41.5d),} Her X-1 (1.7d),
OAO 1657-415 (10.4d) and Vela X-1 (9.0d).
However, modulations at the known orbital periods of Sco X-1 (0.79d) and
the galactic superluminal jet source GRO J1655-40 (2.6d) are absent.

Eclipses of the hard X-ray emission regions are evident in 4U 1700-377, Cen X-3,
Her X-1, \mbox{OAO 1657-415} and Vela X-1.   The broad ingress to eclipse in the Her X-1
light curve is likely produced through the superposition of various pre-eclipse absorption
features which are known to viewed through column densities reaching up to $\sim$ $10^{24}$ $\rm{cm^{-2}}$ (Choi 1993).
Orbital X-ray modulations in Cyg X-1, detected at lower energies (see Priedhorsky, Brandt \& Lund 1995
and references therein), have also been suspected of
being produced by variable absorption.  The sine-like shape and $\approx$5 $\%$ full amplitude modulation
 of the 45-140 keV light curve
for Cyg X-1 (see Figure 3), if due to absorption, would require variations in column
density of order $10^{25}$ $\rm{cm^{-2}}$ lasting over half of its orbit.  This finding, together
with the recent observations of the RXTE/ASM showing the 3-12 keV orbital modulation to
be more prominent during hard state observations (Zhang, Robinson \& Cui 1996), argues
against absorption as
the primary cause for orbital X-ray variability in this source.

\begin{figure*}[t]
\hbox{
\psfig{file=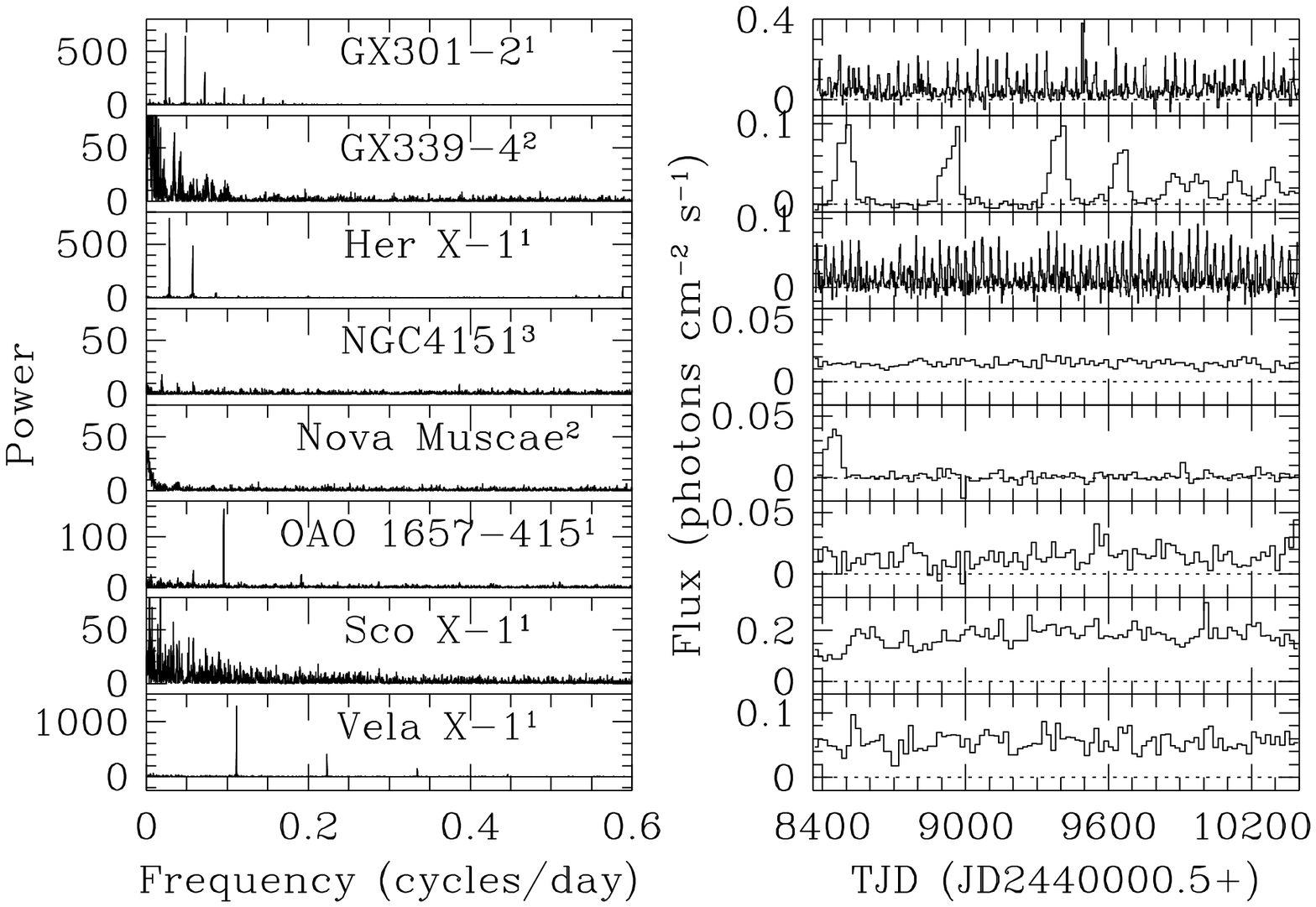,width=6.5in} }
\vspace{-10cm}
\caption{ See caption to Figure 1.  The only distinction is that the light curves for GX 301-2 and Her X-1 are binned in
2 day intervals instead of 20 day intervals to show the evolution of their orbital and precessional modulations, respectively.}
\vspace{-11.5cm}

\hbox{
\psfig{file=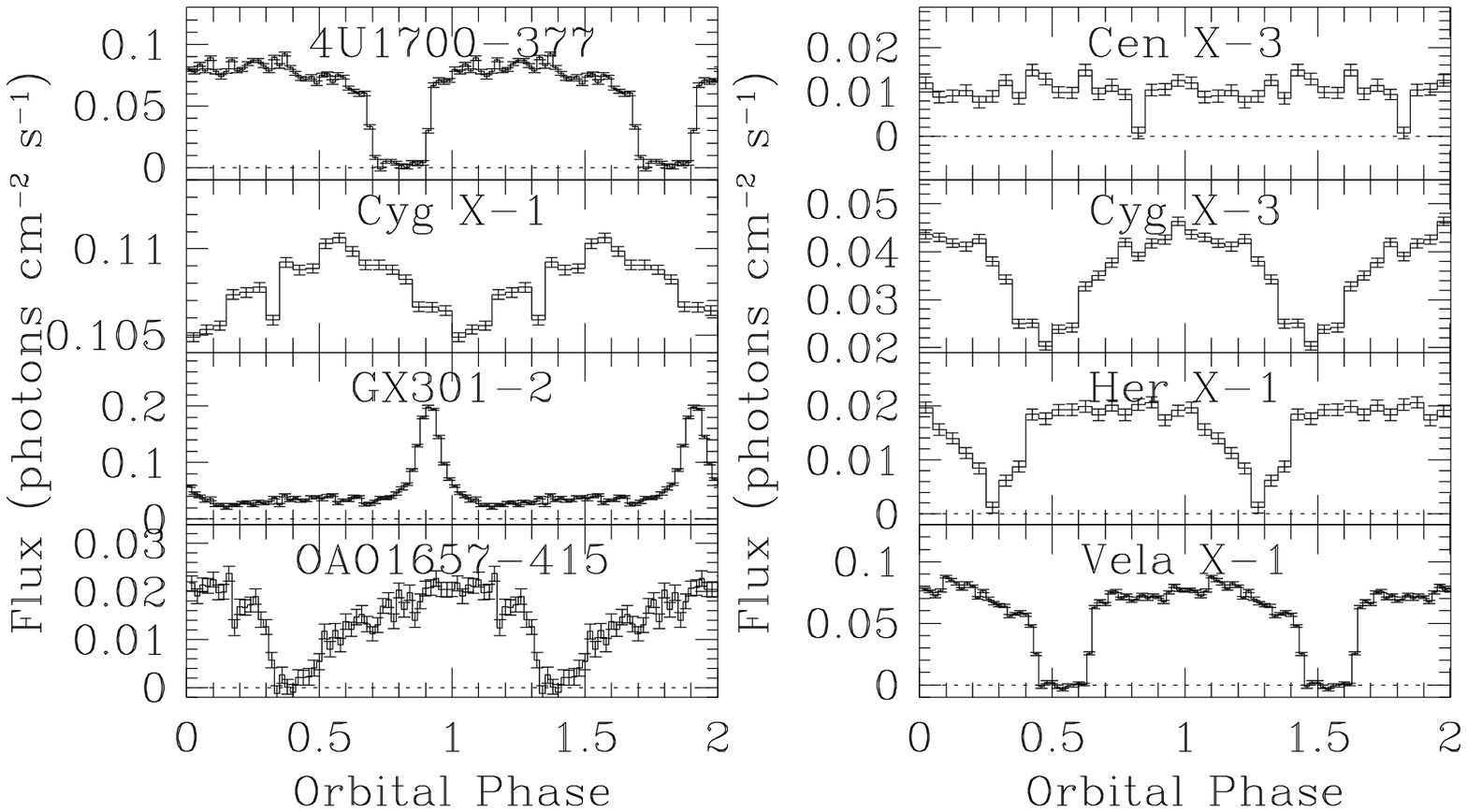,width=6.5in}
     }
\caption{BATSE light curves for the eight binaries showing eclipses or strong orbital modulations are shown folded at
their orbital periods (arbitrary phase zero point).  Error bars correspond
to statistical errors only.}
\end{figure*}

\subsection{Superorbital Modulations and Long-term Variability}

Superorbital X-ray modulations have been detected in several X-ray binaries with the
modulation in \mbox{Her X-1} being the best studied.  However, the physical mechanisms
responsible in Her X-1 and other systems, likely involving a precessing, twisted disk, still require
further study.  In addition, long-term
modulations in these superorbital cycles have not been well-explored.   A total of 57 cycles have been observed
of the 35-day superorbital modulation in Her X-1.  Two peaks per cycle (the so-called ``main high''
and ``short high'' states) are detected in most, but not all, of the 57 cycles.  The short high peaks
are most prominent in those cycles where the main high peaks are bright.  Variations
of more than a factor of two exist in the main high peak flux and its FWHM
on time scales as short 
as two consecutive cycles.  However, there also exist long stretches where the main high peaks
appear bright and broad while in other stretches weaker and more narrow peaks occur.   Transitions between these
states appear as broad modulations in the main high height, as may be seen in Figure 2.  The time scale
for a complete cycle of these transitions ranges from 200 to 700 days and presumably arise
from variations in the mass transfer rate onto the neutron star.

Periodic long-term features in X-ray binaries require several repeated
cycles before they can be confirmed.  For example, in GX 339-4,
BATSE initially observed three strong outbursts with near equal intensities
and near equal separations leading speculation that this initial characteristic outburst time scale
was a fixed, rigid feature in the system.
However, subsequent outbursts clearly showed the time between outbursts
was variable with a significant correlation between outburst fluence and time
since the last outburst.

The variable sources monitored by BATSE are not limited only to
Galactic objects.  We note that long-term variability in 3C273, variability
from weeks to years in Cen A, and the near constant flux from
NGC 4151 are all easily observable by BATSE.
A more comprehensive analysis of the present data set and the further inclusion of weaker sources
in the BATSE database are in progress.

\clearpage


%
%
\end{document}